\documentclass[prd,aps, tightenlines, preprint, preprintnumbers, showpacs, nofootinbib,superscriptaddress]{revtex4}

\usepackage{graphicx}
\usepackage{amsmath,amsthm,amssymb}
\usepackage{mathtools}
\usepackage[usenames,dvipsnames]{color}

\begin{document}
\title{Resumming  subleading Sudakov logarithms in saturation regime }

\author{Jian Zhou}
\affiliation{\normalsize\it  Key Laboratory of Particle Physics and
Particle Irradiation (MOE), Institute of frontier and interdisciplinary science, Shandong
University,  QingDao, Shandong 266237, China}

\begin{abstract}
We investigate the scale dependence of transverse momentum
dependent(TMD) gluon distribution in saturation regime.  We found
that in the Collins-2011 scheme, the scale dependence of small $x$
gluon TMD is governed by the same renormalization group(RG) equation
that holds at moderate or large $x$. Following the standard
procedure, one then can resum both double leading logarithm and
single leading logarithm in saturation regime by jointly solving the
Collins-Soper equation and the RG equation.

\end{abstract}

 \maketitle

\section{Introduction}
One of the central scientific goals to be achieved at the current and future facilities,
including JLab 12 Gev upgrade, RHIC and the planned electron-ion collider(EIC) is to
 reveal  the three-dimensional structure of nucleon/nuclei by measuring final state produced particle
 transverse momentum spectrum in high energy scatterings.
 The extraction of parton transverse momentum dependent(TMD) distributions that encode
information on the internal structure of nucleon/nuclei from physical observables relies on the QCD factorization theorem.
As a leading power approximation, TMD factorization in moderate or large $x$ region has been well established during
the past few decades~\cite{Collins:1981uw,Collins:1981uk,Collins:1984kg,collins}.
 However, since high twist contributions arise from multiple gluon re-scattering
is no longer negligible at small $x$, it is nontrivial to justify TMD factorization in saturation regime.
Some recent attempts to address this issue have been made in
 Refs.~\cite{Dominguez:2010xd,Dominguez:2011wm,Mueller:2012uf,Balitsky:2015qba,Marzani:2015oyb,Zhou:2016tfe,Xiao:2017yya}

The purpose of the current work is to further extend and refine the previous analysis presented in Refs.~\cite{Zhou:2016tfe,Xiao:2017yya}.
The key idea of unifying small $x$ formalism and TMD approach is a two step evolution procedure, which
can be best demonstrated using a color neutral scalar particle production through gluon fusion($gg \rightarrow H
$) process as an example. Below the produced particle mass and transverse momentum
are denoted as $Q$ and $k_\perp$ respectively. For a comparison, we first review the conventional treatment in  moderate $x$ region,
where the calculation of the differential cross section can be formulated in collinear factorization. At high order,
the collinear divergence and various large logarithm terms show up. After absorbing the collinear divergence and the associated
logarithm ${\rm ln}\frac{Q^2}{\mu^2}$ into the renormalized gluon PDF, we are still left with
large double logarithm term $\alpha_s^n{\rm ln}^{2n}\frac{Q^2}{k_\perp^2}$. To facilitate resumming these large logarithms to all orders,
one can introduce gluon TMD distribution.  The large $k_\perp$ logarithms
then can be resummed by solving the Collins-Soper evolution equation~\cite{Collins:1981uw,collins} that governs ${\rm ln}\zeta_c$ dependence of gluon TMD.
Here $\zeta_c$ is a parameter introduced in the Collins 2011 scheme~\cite{collins} for regulating the light cone divergence.
It plays a role of varying hard scale which allows one to smoothly evolve from the scale $Q^2$ down to $k_\perp^2$.

When the center mass of energy $S$ is much larger than $Q^2$, the large logarithm
${\rm ln}\frac{S}{Q^2} \sim {\rm ln}\frac{1}{x}$ appears in high order calculation could be more important than
 the logarithm ${\rm ln}\frac{Q^2}{\mu^2}$.
One thus should formulate the calculation in
color glass condensate(CGC) effective theory~\cite{McLerran:1993ni,JalilianMarian:1997jx}
to first take care of the logarithm
 ${\rm ln}\frac{S}{Q^2}$. They can be summed by solving the
 Balitsky-Kovchegov(BK)~\cite{Balitsky:1995ub,Kovchegov:1999yj}
  equation that  describes ${\rm ln}\frac{1}{x}$ dependence of  multiple point function---the basic nonperturbative
 ingredient in CGC calculation.
 In the kinematic region where $Q^2 \gg k_\perp^2$,
 logarithm $\alpha_s^n{\rm ln}^{2n}\frac{Q^2}{k_\perp^2}$ terms also arise in high  order calculations.
 When these logarithms are much larger than leading order
 but high twist contributions suppressed by the power of $\frac{k_\perp^2}{Q^2}$~\cite{Mueller:2012uf},
 TMD factorization should be employed
 where all subleading power contributions are systematically ignored. The equivalence between
the leading power part of the CGC result and TMD factorization calculation at tree level can be
verified by utilizing the operator relation between the derivative of multiple point function
 and gluon TMD matrix element~\cite{Dominguez:2010xd,Dominguez:2011wm}.
 What we gain by making the leading power approximation is that
 large $\alpha_s^n{\rm ln}^{2n}\frac{Q^2}{k_\perp^2}$ logarithm terms
 can be resummed to all orders in the context of TMD factorization.

 To achieve such a resummation, it is necessary to show that the
 properly defined gluon TMD  accommodates the similar large  logarithm
 and satisfies the Collins-Soper equation in the small $x$ limit.
 It has indeed been verified in a recent work~\cite{Zhou:2016tfe} that
 gluon TMD computed at the next to leading order in a quark target model satisfies both the
 Balitsky-Fadin-Kuraev-Lipatov(BFKL) equation~\cite{Kuraev:1977fs,Balitsky:1978ic} and the
 Collins-Soper equation in the small $x$ limit.
 The similar analysis  was later extended to saturation regime by
 calculating small $x$ gluon TMD in terms of multiple point functions using CGC approach~\cite{Xiao:2017yya}.
 Schematically,
 the derived Weizs\"{a}cker-Williams (WW) type gluon TMD takes the following form,
 \begin{eqnarray}
xG(x,k_\perp,\zeta_c^2=\mu_F^2=Q^2)&=&-\frac{2}{\alpha_s}
\int \frac{d^2x_\perp d^2y_\perp}{(2\pi)^2}e^{ik_\perp\cdot b_\perp}
\nonumber \\ &\times& {\cal H}(\alpha_s(Q))
\text {Exp}\left \{ -\int_{\mu_b}^Q \frac{d\mu}{\mu}\left ( A \
{\rm ln}\frac{Q^2}{\mu^2}+B \right ) \right \}
 {\cal F}(x_\perp,y_\perp)
\end{eqnarray}
where  $\mu_b=\frac{2e^{-\gamma_E}}{|b_\perp|}$ with $b_\perp=x_\perp-y_\perp$,
and  $\mu_F$ is the factorization scale.
 ${\cal F}$ is the Fourier transform of the WW gluon distribution,
\begin{equation}
{\cal F}(x_\perp,y_\perp)= \left\langle{\rm Tr}\left[
\partial_\perp^i U(x_\perp) U^\dagger(y_\perp)\partial_\perp^i
U(y_\perp)U^\dagger(x_\perp)\right]\right\rangle \ \label{TMD0}
\end{equation}
where $U(x_\perp)$ is the Wilson line in the fundamental representation.
It absorbs all large logarithm ${\rm ln}\frac{1}{x}$  terms from the hard part with the help of the BK equation.
The remaining logarithms are resummed into the exponentiation known as the Sudakov factor by solving the Collins-Soper equation.  We are
eventually left with a hard coefficient ${\cal H}(\alpha_s(Q))$ that only has finite contributions.
The  Similar result holds for the dipole-type gluon distribution.

 The hard coefficients $A$ and $B$ can be calculated perturbatively. In the previous work~\cite{Xiao:2017yya},
 we only took into account the double leading logarithm contribution in CGC framework,
 and determined the coefficient $A$
to be  $A=\frac{\alpha_sC_A}{\pi}$ at leading order,
 which is the same as the one in the standard Collins-Soper-Sterman(CSS) formalism~\cite{Collins:1984kg}.
The purpose of the current work is to sort out the single leading logarithm contribution in saturation regime,
i.e. fixing the coefficient $B$.
To this end, one has to  study not only the ${\rm ln}\zeta_c$ dependence but also the factorization
scale $\mu_F$ dependence of small $x$ gluon TMD. In other words, we aim at deriving a renormalization group(RG)
 equation for the gluon TMD in saturation regime.  By jointly solving  the RG equation and the Collins-Soper equation,
 one is  able to resum both the double leading logarithm and single leading logarithm contributions.

 From a theoretical point of view, completing the previous analysis on $k_\perp$ resummation in
 saturation regime is interesting in its own right.  On the other hand, the present work is further  motivated by the fact that
 very rich polarization dependent phenomenology in saturation regime
  has been discovered in recent
   years~\cite{Metz:2011wb,Zhou:2013gsa,Boer:2015pni,Boer:2016xqr,Dong:2018wsp,Kovchegov:2015pbl,Hatta:2016dxp,Zhou:2016rnt,Boer:2018vdi}.
 It is time to lay down a solid theoretical ground for performing phenomenological studies of the
 relevant physical observables which can be measured at RHIC, LHC, and the planned EIC.

 The rest of this paper is organized as follows. In the next section, we compute the
anomalous dimension of small $x$ gluon TMD by isolating the ultraviolet(UV) divergent part. The most
nontrivial part of our analysis  is to investigate how the UV divergence is affected
in the  presence  of multiple gluon re-scattering.   The detailed derivation will be presented. The paper
is summarized in Sec.III.

\section{Derivation}
There are two widely used $k_\perp$ dependent unpolarized gluon distributions with different gauge link
structures: (1) the WW type distribution with a staple like gauge link, and (2) the dipole type distribution
with a close loop gauge link.  These two type gluon distributions can be directly probed through two-particle
correlation in different high energy scattering
reactions~\cite{Dominguez:2010xd,Dominguez:2011wm,Akcakaya:2012si,Kotko:2015ura,Boer:2017xpy,Benic:2017znu}.
 In Ref.~\cite{Xiao:2017yya}, we demonstrated that
both gluon TMDs computed in CGC formalism satisfy the Collins-Soper equation after matching them onto the
renormalized quadrupole and dipole amplitudes respectively.
At leading order, the Collins-Soper equation reads~\cite{collins},
\begin{eqnarray}
\frac{\partial ~{\rm ln}G(x,b_\perp,\mu^2,\zeta_c^2)}{\partial ~{\rm ln}\zeta_c}
=K(b_\perp,\mu) =-\frac{\alpha_s C_A}{\pi} {\rm ln}\frac{\mu^2 b_\perp^2}{4 e^{-2 \gamma_E}}
\end{eqnarray}
where $G(x,b_\perp,\mu^2,\zeta_c^2)$ is the Fourier transform of gluon TMD,
which can be related to the derivative of the quadrupole amplitude for the WW case.
 The logarithm ${\rm ln}\frac{1}{x}$ dependence of the operator $G(x,b_\perp,\mu^2,\zeta_c^2)$ is
 described by the BK equation.
On the other hand, the factorization scale dependence of gluon TMD is governed  by the renormalization group
equation which takes form,
\begin{eqnarray}
\frac{ {\rm d ~ln}G(x,b_\perp,\mu^2,\zeta_c^2)}{ {\rm d~ ln}\mu}=\gamma_G \left(g(\mu),\zeta_c^2/\mu^2 \right )
\end{eqnarray}
The anomalous dimension $\gamma_G$ is also the function of $\zeta_c$. Its $\zeta_c$ dependence can be explicitly
separated out as the following~\cite{collins},
\begin{eqnarray}
\gamma_G \left(g(\mu),\zeta_c^2/\mu^2 \right )&=&\gamma_G \left(g(\mu),1 \right )-
\frac{1}{2} \gamma_K(g(\mu)) {\rm ln}\frac{\zeta_c^2}{\mu^2}
\end{eqnarray}
with $ \gamma_K(g(\mu)=-\frac{{\rm d }K(b_\perp, \mu)}{ {\rm d~ln}\mu}=\frac{2\alpha_s C_A}{\pi}$ at one loop order.
Following the standard procedure, it can be readily deduced from the evolution equations
\footnote{It has been found~\cite{Scimemi:2018xaf} that the final implementation of the TMD
evolution  depends on the particular choice of integration path in the ($\mu$, $\zeta_c$) plane.
This deserves further investigation. },
\begin{eqnarray}
G(x,b_\perp,Q^2,Q^2)=\text {Exp} \left \{ \int_{\mu_b}^Q
\frac{d \mu}{\mu}  \left (\gamma_G \left(g(\mu),1 \right )-
\frac{\alpha_s C_A}{\pi} {\rm ln} \frac{Q^2}{\mu^2} \right  ) \right \}
 G(x,b_\perp,\mu_b^2,\mu_b^2)
\end{eqnarray}
where the large logarithm is resummed into the Sudakov factor.

In the dilute limit, it is shown~\cite{Zhou:2016tfe} that both the double leading and single leading logarithms can be resummed.
In the saturation regime, is only the double leading logarithm contribution
$\text {Exp} \left \{ -\int_{\mu_b}^Q \frac{d \mu}{\mu}
\frac{\alpha_s C_A}{\pi} {\rm ln} \frac{Q^2}{\mu^2}  \right \}$ took into account in the
previous analysis~\cite{Xiao:2017yya}.
The value of $\gamma_G \left(g(\mu),1 \right )$ is not yet fixed for the saturation case.
The purpose of the present work is to compute the anomalous dimension of small $x$
gluon TMD in saturation regime. Once the anomalous dimension is worked out, the single leading logarithm
can be resummed to all orders by solving the CS and RG equations as shown above.

 As an example, we focus on the WW case in this paper.  The calculation can be
straightforwardly extended to the dipole case.
Our starting point is the the matrix element definition of the WW gluon distribution,
\begin{eqnarray}
xG(x,k_\perp)\!=\!\!\int \!\! \frac{d\xi^- d^2\xi_\perp}{(2\pi)^3P^+}e^{ik_\perp \cdot \xi_\perp-ixP^+\xi^-}
\langle P | F^{+i}_a(\xi^-,\xi_\perp) {\cal L}_{ab}^\dag(\xi^-,\xi_\perp) {\cal L}_{bc}(0,0_\perp)
 F^{+i}_c(0) | P \rangle \label{TMDa}
\end{eqnarray}
where $F^{+i}_a(\xi^-,\xi_\perp)$ is the gauge field strength tensor and
the gauge link is further fixed to be the past pointing one in the adjoint representation,
\begin{eqnarray}
{\cal L}_{ab}^\dag(\xi^-,\xi_\perp)&=&{\cal P}
{\rm exp} \left [ -g_0 \int_{-\infty}^{\xi^-} dz^- f^{dab}A^+_d(z^-,\ \xi_\perp) \right ]
\\
{\cal L}_{bc}(0,0_\perp)&=&{\cal P}
{\rm exp} \left [ -g_0 \int^{-\infty}_{0} dz^- f^{dbc}A^+_d(z^-,\ \xi_\perp) \right ]
\end{eqnarray}
The WW type gluon TMD also can be defined in the fundamental representation,
\begin{eqnarray}
xG(x,k_\perp)\!=\!\! 2\int \!\! \frac{d\xi^- d^2\xi_\perp}{(2\pi)^3P^+}e^{ik_\perp \cdot \xi_\perp-ixP^+\xi^-}
\langle P | {\rm Tr} \left [
 F^{+i}_a(\xi^-,\xi_\perp) U^{[+]\dag}
 F^{+i}_c(0)  U^{[+]}\right ] | P \rangle \label{TMDf}
\end{eqnarray}
where $U$ denotes the gauge link in the fundamental representation.

One can readily  determine the anomalous dimension by isolating the ultraviolet(UV) divergence of small $x$
gluon TMD. To do so, one has to go beyond the conventional treatment of  small $x$ formalism in which the Eikonal approximation
is applied everywhere.  This is because the UV divergent part does not have $1/x$ enhancement and  could
be missed in the leading power small $x$ approximation. We thus have to carry out the calculation in the full QCD.
The extra care should be taken when performing the Eikonal approximation to simplify calculation.
\begin{figure}[t]
\begin{center}
\includegraphics[width=9 cm]{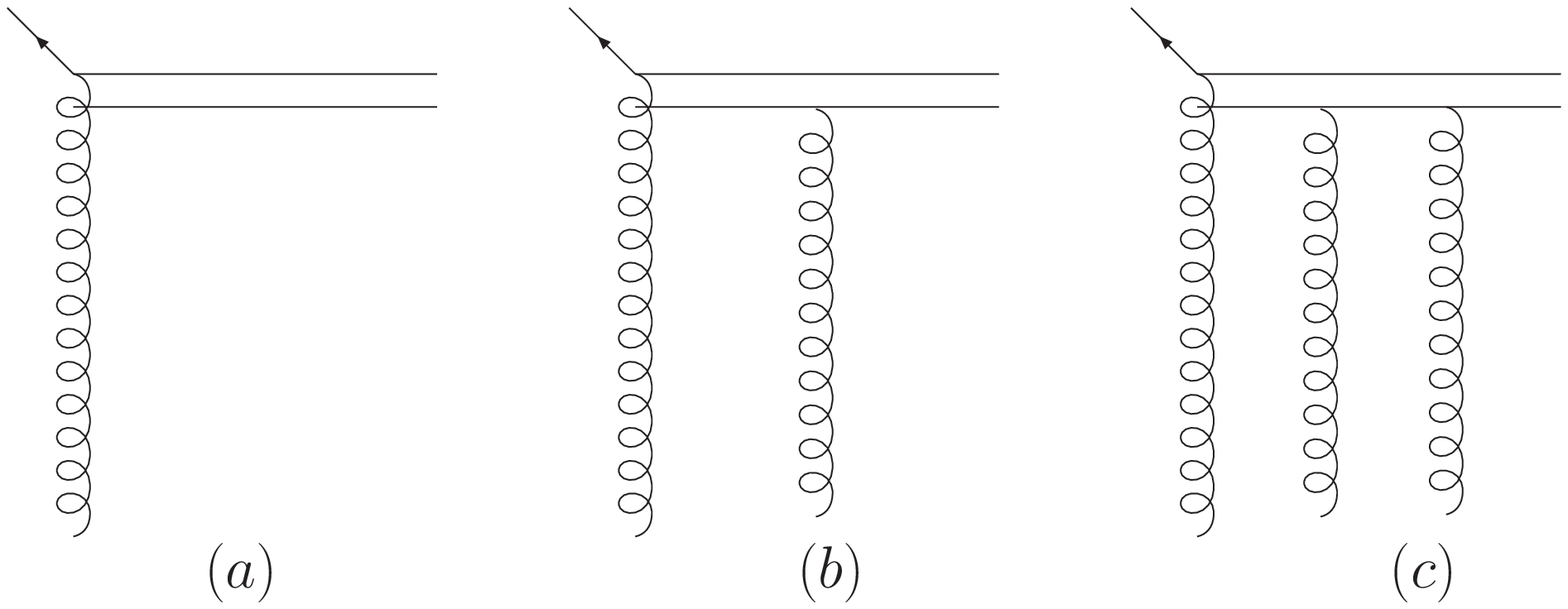}
\caption[] {Tree diagrams contributing to the WW type gluon TMD.  The double lines represent gauge link
in the adjoint representation. }
 \label{1}
\end{center}
\end{figure}

In order to fix conventions and to do a warm up exercise,
we start with the tree level calculation, though the UV divergence is absent at the tree level.
 Diagrams illustrated in Fig.1 give rise to the leading order contributions. In the small
 $x$ limit, the dominant contribution is from the $A^+$ component. It is trivial
 to compute the graph Fig.1(a) and its conjugate part,  which lead to,
\begin{eqnarray}
xG_0(x,k_\perp)\!=\! 4 \pi \int \!\! \frac{dx^-d^2x_\perp}{(2\pi)^2} \frac{dy^-d^2 y_\perp}{(2\pi)^2}
e^{ik_\perp \cdot(x_\perp-y_\perp)} \langle
  \partial_i A^+_a(x^-,x_\perp)\partial_i A^+_a(y^-,y_\perp) \rangle_x
\end{eqnarray}
It is well known that the gauge link is built through gluon re-scattering. The diagram Fig.1(b) gives rise to the first nontrivial
term of the Taylor expansion of the gauge link,
\begin{eqnarray}
xG_1(x,k_\perp)\!&=&\! 4 \pi \int \!\! \frac{dx^-d^2x_\perp}{(2\pi)^2} \frac{dy^-d^2 y_\perp}{(2\pi)^2}
e^{ik_\perp \cdot(x_\perp-y_\perp)}
\nonumber \\ &&\times \langle
  \partial_i A^+_a(x^-,x_\perp) \left (- g_0\!\int_{y^-}^{-\infty}\!\!\!\! dz^- f^{bac}A^+_b(z^-,y_\perp) \right ) \partial_i A^+_c(y^-,y_\perp) \rangle_x
\end{eqnarray}
where $g_0$ associated with the gauge potential $A^+_b(z^-,y_\perp)$ is the bare strong coupling constant
, which will be renormalized after including one loop correction.
Similarly, the graph Fig.1(c) results in,
\begin{eqnarray}
xG_2(x,k_\perp)\!&=&\! 4 \pi \int \!\!\! \frac{dx^-d^2x_\perp}{(2\pi)^2} \frac{dy^-d^2 y_\perp}{(2\pi)^2}
e^{ik_\perp \cdot(x_\perp-y_\perp)} \langle
  \partial_i A^+_a(x^-,x_\perp)
  \nonumber \\&&\times
  \left (g_0^2 \!\int_{y^-}^{-\infty}\!\!\!\! dz^- f^{dab}A^+_d(z^-,y_\perp)
  \!\int_{y^-}^{z^-}\!\!\!\! dz_1^- f^{bdc}A^+_b(z_1^-,y_\perp)\right )
  \partial_i A^+_c(y^-,y_\perp) \rangle_x
\end{eqnarray}
It is straightforward to resum gluon re-scattering to all orders. The WW type gluon TMD in the small $x$
limit eventually can be cast into the following form in the fundamental representation,
\begin{eqnarray}
xG(x,k_\perp)\!=\!\! -\frac{8 \pi}{g_0^2} \int \!\! \frac{d^2x_\perp}{(2\pi)^2} \frac{d^2 y_\perp}{(2\pi)^2}
e^{ik_\perp \cdot(x_\perp-y_\perp)} \langle {\rm Tr}
 [ \partial_i U(x_\perp)]U^\dag(y_\perp) [\partial_i U(y_\perp)] U^\dag(x_\perp) \rangle_x \label{TMDx}
\end{eqnarray}
where the strong coupling constant appear in the Wilson lines is the bare one.
The above matrix element obtained through tree diagram calculation
 is consistent with the matrix element definition given in Eq.\ref{TMD0},
 which only captures the leading contribution in the power of $1/x$. In contrast, the gluon TMD definition
 Eq.\ref{TMDa}(or Eq.\ref{TMDf})  is valid at arbitrary  $x$, and keeps not only the leading ${\rm ln}\frac{1}{x}$
 terms but also the leading contributions
 in the power of $\mu^2/\zeta_c^2$ and  $k_\perp^2/\zeta_c^2$
 .  Therefore, the correct UV behavior and the
 ${\rm ln} \zeta_c$ dependence of gluon TMD only can be obtained by computing the expectation value of
 the matrix element in Eq.\ref{TMDa}(or Eq.\ref{TMDf}), rather than the matrix element given in Eq.\ref{TMDx}.

UV divergence only arises in virtual corrections.
There are four virtual graphs without gluon re-scattering as shown in Fig.2.
 It is easy to verify that the contribution from the graph Fig.2(d) vanishes.
We now  start with computing the vertex correction shown in Fig.2(a).
To avoid the interaction between the radiated gluon and
color source inside target, our calculation is performed in the light cone gauge ($A^-=0$), in which gluon propagator reads,
\begin{eqnarray}
\left ( -g^{\mu \nu} +\frac{p^\mu l^\nu+p^\nu l^\mu}{l \cdot p-i\epsilon}\right ) \frac{i}{l^2+i\epsilon}
\end{eqnarray}
where the prescription $\frac{1}{l \cdot p-i\epsilon}$ for regulating the light cone divergence is proven to
be the most convenient choice for our calculation. The contribution from the graph Fig.2(a) is expressed as the product of the
corresponding hard part and the gluon TMD matrix element without gauge link being included,
\begin{eqnarray}
\text{Fig.2(a)} \propto {\cal H}_{2a}(k)
xG_0(x,k_\perp)
\end{eqnarray}
The hard part is given by,
\begin{eqnarray}
{\cal H}_{2a}(k)  \!  = \!\frac{\alpha_s C_A}{4 \pi^3}\!\!  \int \!\!dl^+ \!d^2l_\perp dl^- \!
\frac{-i\left [(k_\perp^2-k_\perp \cdot l_\perp)(l_\perp^2-k_\perp \cdot
l_\perp) -l^-k^+ k_\perp \cdot l_\perp \right ]}{\left[ 2l^+l^- \!-\!l_\perp^2\!+\!i
\epsilon \right ]\left[ 2l^-(l^+\!\!-\!k^+)\!-\!(l_\perp\!-\!k_\perp)^2 \!+\!i
\epsilon \right ]\left [l^-\!\!-\!i\epsilon \right ]\left [l^+\!+\!i\epsilon
\right ]k_\perp^2}
\end{eqnarray}
where $k_\perp^2$ in the denominator arises when we make the following conversion by partial integration,
\begin{eqnarray}
k_\perp^2 \langle    A^+_a(x^-,x_\perp) A^+_a(y^-,y_\perp) \rangle
\Longrightarrow \langle
  \partial_i A^+_a(x^-,x_\perp)\partial_i A^+_a(y^-,y_\perp) \rangle
\end{eqnarray}
We proceed by performing contour integration on $l^-$,
\begin{eqnarray}
{\cal H}_{2a}(k)  &=&
  -\frac{\alpha_s C_A}{ 2\pi^2}\int d^2l_\perp \int_0^{k^+} dl^+
  \frac{(k_\perp^2-k_\perp \cdot l_\perp)(l_\perp^2-k_\perp \cdot
l_\perp) -l_\perp^2 k_\perp \cdot l_\perp k^+/(2l^+)}
{\left[ (l^+-k^+)l_\perp^2-(l_\perp-k_\perp)^2 l^+ +i\epsilon \right ]l_\perp^2k_\perp^2}
\nonumber \\&&+
 \frac{\alpha_s C_A}{ 2\pi^2} \int d^2l_\perp \int_{k^+}^\infty dl^+
  \frac{(k_\perp^2-k_\perp \cdot l_\perp)(l_\perp^2-k_\perp \cdot
l_\perp) }
{(l_\perp-k_\perp)^2 l_\perp^2 l^+k_\perp^2}
\end{eqnarray}
As we are only interested in the UV behavior of the small $x$ gluon  TMD, the external transverse momentum $k_\perp$ can be neglected.
It is then  trivial to carry out the elementary integration  for $l^+$. One arrives at,
\begin{figure}[t]
\begin{center}
\includegraphics[width=12 cm]{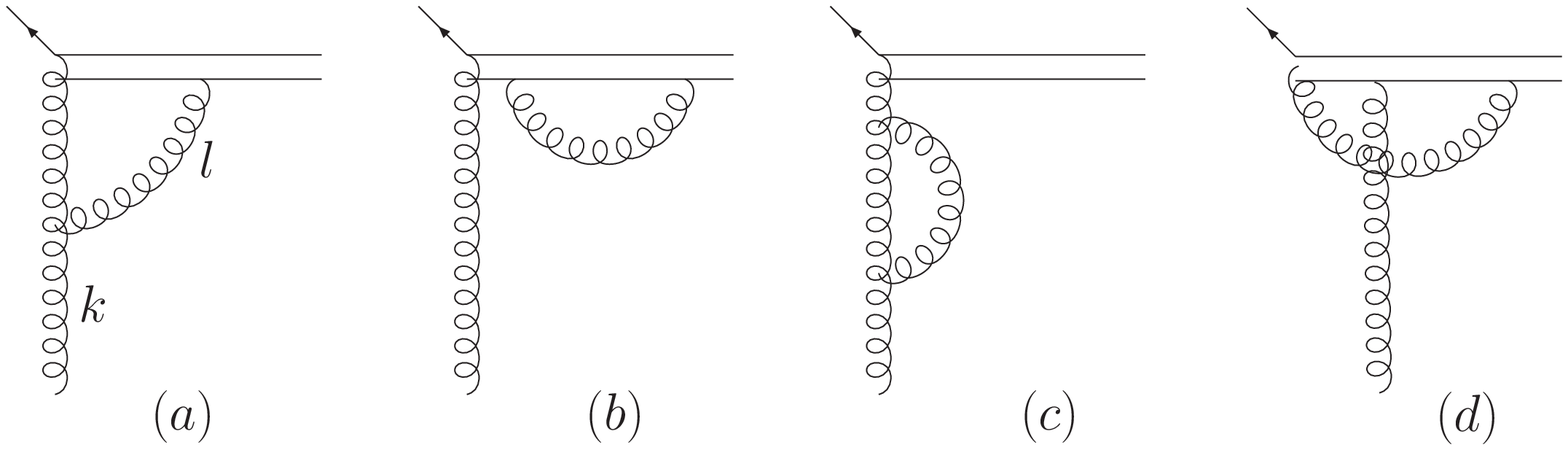}
\caption[] { One loop virtual correction to the gluon TMD without taking into account gauge link.
In the small $x$ limit, the $A^+$ component of the incoming gluon gives rise to the dominant contribution.
The diagram d has vanishing contribution. The ghost contribution is absent in the gauge $A^-=0$. }
 \label{1}
\end{center}
\end{figure}
\begin{eqnarray}
{\cal H}_{2a}(k)  &\approx&
  -\frac{\alpha_s C_A}{2 \pi^2} \int d^2l_\perp \int_0^{k^+} dl^+
   \frac{l^+-k^+}{l_\perp^2}
+ \frac{\alpha_s C_A}{2 \pi^2} \int d^2l_\perp  \int_{k^+}^\infty dl^+  \frac{1}{l^+ l_\perp^2}
\nonumber \\&=&\frac{\alpha_s C_A}{2 \pi^2}  \int \frac{d^2l_\perp}{l_\perp^2}
 \left \{ \frac{1}{2}+ \int_{k^+}^\infty \frac{dl^+}{l^+} \right \}
\end{eqnarray}
In contrast to a covariant gauge calculation, the vertex correction we obtained is free from the Collins-Soper type light cone divergence.
But the second integration $ \int_{k^+}^\infty \frac{dl^+}{l^+}$ has the BFKL/BK type light cone divergence when $l^+$ goes infinity,
and leads to the small $x$ evolution of gluon TMD, that is beyond the scope of the current work.

We turn to discuss the Wilson line self energy correction,
\begin{eqnarray}
\text{Fig.2(b)} \propto {\cal H}_{2b}(k) \
xG_0(x,k_\perp)
\end{eqnarray}
with the hard part,
\begin{eqnarray}
{\cal H}_{2b}(k)&= &\frac{\alpha_s C_A}{4 \pi^3}  \int dl^+ d^2l_\perp dl^-
\frac{i}{\left[ 2l^+l^- -l_\perp^2+i
\epsilon \right ]\left [l^--i\epsilon \right ]\left [ l^++i\epsilon \right] }
 \nonumber \\&=&
 - \frac{\alpha_s C_A}{ 2\pi^2} \int \frac{d^2l_\perp}{l_\perp^2} \int_0^{\infty} \frac{dl^+}{l^+}
\end{eqnarray}
which contains the light cone divergence when $l^+$ goes to zero. Such end point singularity can be cured
by introducing a soft factor in the Collins-2011 scheme.

The gluon self energy graph Fig.2(c) gives,
\begin{eqnarray}
\text{Fig.2(c)} \propto \frac{1}{2} {\cal H}_{2c}(k) \
xG_0(x,k_\perp)
\end{eqnarray}
According to the LSZ reduction formula,  half the one loop correction of the
gluon propagator contributes to the anomalous dimension of the gluon TMD, while another half contributes to
the renormalization of gauge field.
That is why we include a factor $\frac{1}{2}$ in the above equation.
Once again, we use the residue theorem to perform the $l^-$ integration in the hard part. This gives,
\begin{eqnarray}
{\cal H}_{2c}(k)&=& \frac{1}{2} \frac{\alpha_s C_A}{4\pi^3} \!\!  \int \!\! dl^+ d^2l_\perp dl^-
\frac{4i \left [(2l^+\!-k^+)l^-\!-k_\perp^2 \right ]}{\left[ 2l^+l^-\! -l_\perp^2+i
\epsilon \right ]\left[ 2l^-(l^+\!-k^+)-(l_\perp\!-k_\perp)^2 +i
\epsilon \right ] k_\perp^2}
 \nonumber \\&=&\frac{1}{2} \frac{\alpha_s C_A}{2 \pi^2} \int d^2l_\perp \int_0^{k^+} dl^+
\frac{2(2l^+-k^+)l_\perp^2/l^+-2k_\perp^2}{2\left[ (l^+-k^+)l_\perp^2-l^+(l_\perp-k_\perp)^2  \right ]k_\perp^2}
\end{eqnarray}
As before, to explicitly isolate the UV pole contribution, the external transverse momentum $k_\perp$ is set to
be zero. The $l^+$ integration can be done by very elementary methods,
\begin{eqnarray}
{\cal H}_{2c}(k)&\approx&
   \frac{1}{2} \frac{\alpha_s C_A}{ 2\pi^2}\int d^2l_\perp \int_0^{k^+} dl^+
  \frac{\left [(k^+)^2+4k^+l^+-4(l^+)^2\right ]}{(k^+)^3 l_\perp^2}
   \nonumber \\&=&
 \frac{\alpha_s C_A}{ 2\pi^2} \int \frac{d^2l_\perp}{l_\perp^2} \frac{5}{6}
\end{eqnarray}
Put all contributions from Fig.2 together,
\begin{eqnarray}
{\cal H}_{2a}(k)+ {\cal H}_{2b}(k) +\frac{1}{2}{\cal H}_{2c}(k) \!&\approx&\! \frac{\alpha_s C_A}{2\pi^2}
  (2\pi \mu)^{2\epsilon}
\!\int \frac{d^{2-2\epsilon} l_\perp}{l_\perp^2}\left \{\frac{11}{12}-\int_0^{k^+}
\frac{dl^+}{l^+} \right \} \label{UV}
\end{eqnarray}
where  dimensional regularization is introduced. Some of finite terms might be missed at intermediate steps.
However, such treatment is sufficient as we only need to compute UV pole terms for the current purpose.
 It is interesting to notice that the UV divergence cancels out in the phase space region
 $k^+ \leq l^+ \leq \infty$. This is consistent with the observation
 that the evolution kernels of the BFKL/BK  equations are UV finite. This is also the precise reason
 why one has to formulate the calculation in full QCD rather than in small $x$ formalism where many
 sub-leading terms in the power of $1/x$ are missed, including UV pole terms.

The end point singularity in the second term in Eq.\ref{UV} is canceled by the soft factor in the Collins-2011 scheme.
Combining with contributions from the hermitian conjugate diagrams, the subtracted gluon TMD then takes form
\begin{eqnarray}
xG(x,k_\perp,\mu^2,\zeta_c^2)\!&\approx&\! \frac{\alpha_s C_A}{2\pi^2}
 (2\pi \mu)^{2\epsilon}
\!\int \frac{d^{2-2\epsilon} l_\perp}{l_\perp^2}\left \{\frac{11}{6}-
{\rm ln}\frac{\zeta_c^2}{l_\perp^2} \right \} xG_0(x,k_\perp)+ \text{UV \ c.t.}
\end{eqnarray}
One finds that both the factorization scale $\mu$ and the parameter $\zeta_c$ dependence of the gluon TMD
 show up at the next to leading order.
The UV counterterm is added in the above equation to give a finite result at $\epsilon=0$.
Note that the collinear divergence in our calculation is absent
 once the incoming gluon transverse momentum $k_\perp$ is restored.
 In a conventional collinear factorization calculation,  the remaining collinear divergence can be removed
after matching TMD onto gluon PDF.

The UV counterterm in the $\overline{\text MS} $ scheme is determined as~\cite{collins},
\begin{eqnarray}
\text{UV \ c.t.}=-\frac{\alpha_s C_A}{2\pi} S_\epsilon \left [\frac{1}{\epsilon^2}+\frac{1}{\epsilon}
\left ( \frac{11}{6}- {\rm ln}\frac{\zeta_c^2}{\mu^2} \right \}
\right ]
\end{eqnarray}
where $S_\epsilon=\frac{(4\pi)^\epsilon}{\Gamma(1-\epsilon)}$.
The anomalous dimension of the gluon TMD can be computed accordingly,
\begin{eqnarray}
\gamma_G \left(g(\mu),\zeta_c^2/\mu^2 \right )=\frac{ {\rm d } \left (-{\frac{\alpha_s C_A}{2\pi} S_\epsilon \left [\frac{1}{\epsilon^2}+\frac{1}{\epsilon}
\left ( \frac{11}{6}- {\rm ln}\frac{\zeta_c^2}{\mu^2} \right \}
\right ]}\right )}{ {\rm d} {\rm ln}\mu}=\frac{\alpha_s C_A}{\pi}
\left ( \frac{11}{6}- {\rm ln}\frac{\zeta_c^2}{\mu^2} \right )
\end{eqnarray}
which is the same as the standard one.  With this anomalous dimension, one reproduces the common Sudakov factor
including both double leading logarithm and single leading logarithm contributions in the dilute limit.
\begin{figure}[t]
\begin{center}
\includegraphics[width=14 cm]{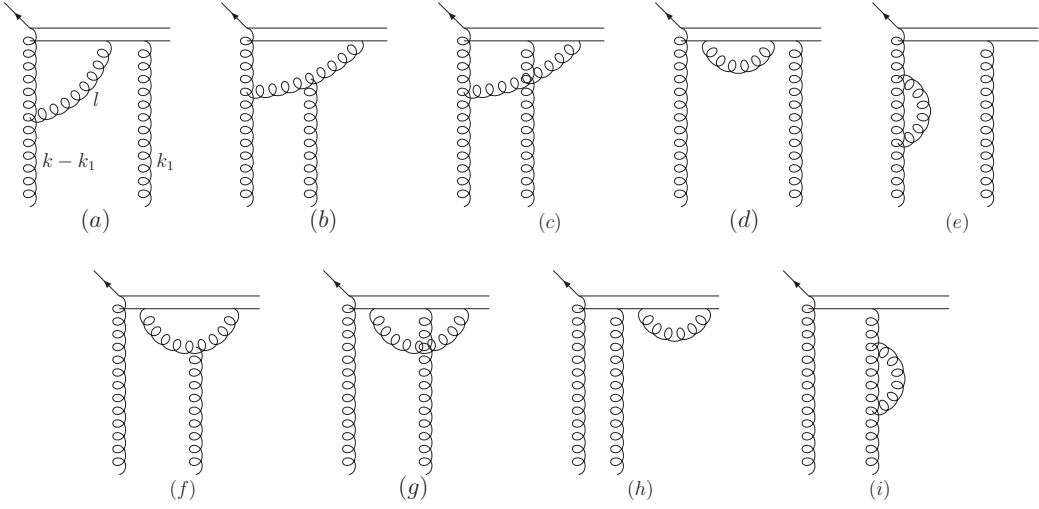}
\caption[] { Virtual correction with one gluon re-scattering.
 Collinear reducible diagrams~\cite{collins} are not shown here.}
 \label{1}
\end{center}
\end{figure}

We now calculate virtual correction in the presence of gluon re-scattering. As argued before,
it is not appropriate to first resum all order gluon re-scattering into the Wilson lines by applying the
Eikonal approximation because the UV divergent part is the sub-leading contribution without $1/x$ enhancement.
Instead, one should work out the UV part before resumming gluon re-scattering. Here we start with one
gluon re-scattering case.

First of all, it is easy to check that diagrams with four gluon vertex have vanishing contribution in the
gauge we specified above.  We start evaluating the vertex correction from
 Fig.3(a), Fig.3(b), and Fig.3(c). It is convenient to calculate the following
combination,
\begin{eqnarray}
&& \!\!\!\!\!\!\!\!\!\!\!\!\!
\frac{1}{2}\text{Fig.3(a)}+\text{Fig.3(c)} \propto
{\cal H}_{\frac{a}{2}+c}(k,k_1)\int
\frac{dx^-d^2x_\perp}{(2\pi)^2} \frac{dy^-d^2 y_\perp}{(2\pi)^2}\frac{dz^-d^2 z_\perp}{(2\pi)^2}
\nonumber \\ && \!\!\!\!\!\!\!\!\! \times
e^{ik_\perp \! \cdot x_\perp}e^{-i(k_\perp\!-k_{1\perp}) \cdot y_\perp}e^{-ik_{1\perp } \! \cdot z_\perp}
e^{i(k^+\!-k_{1}^+)   y^-}\!e^{ik_{1}^+   z^-} \!
\langle  A^+_a(x^-\!,x_\perp)  g_0f^{bac}A^+_b(z^-\!,z_\perp)  A^+_c(y^-\!,y_\perp) \rangle_x
\end{eqnarray}
with the hard part being given by,
\begin{eqnarray}
&& \!\!\!\!\!\!\!\!\!\!\!\!\!
{\cal H}_{\frac{a}{2}+c}(k,k_1) \propto \frac{C_A}{2} \frac{\alpha_s}{4 \pi^3}
\int d^2k_{1\perp} d k_1^+
\int dl^+ d^2l_\perp dl^- \frac{-i}{k_1^+ + i \epsilon}
\nonumber \\&\times&\!\!
\frac{[k_\perp \cdot(k_\perp-k_{1\perp}-l_\perp)][l_\perp \cdot (l_\perp-k_\perp+k_{1\perp})
] -l^-k^+ k_\perp \cdot l_\perp}{\left[ 2l^+l^- -l_\perp^2+i
\epsilon \right ]\left[ 2l^-(l^+-k^++k^+_1)-(l_\perp-k_\perp+k_{1\perp})^2 +i
\epsilon \right ]\left [l^--i\epsilon \right ]\left [l^++i\epsilon\right ]}
\end{eqnarray}
As stated previously, we do not aim at getting the complete result.
To clearly extract the UV divergent part associated with the leading power contribution,
we set $k_{1\perp}$ to be zero and make the Taylor expansion in terms of the power
 $k_\perp/l_\perp$,
\begin{eqnarray}
&& \!\!\!\!\!\!\!\!\!\!\!\!\!
{\cal H}_{\frac{a}{2}+c}(k,k_1)  \propto \frac{C_A}{2} \frac{\alpha_s}{2 \pi^2}
\int d^2k_{1\perp} d k_1^+ \int d^2l_\perp
 \frac{-1}{k_1^+ + i \epsilon}
\nonumber \\&\times&
  \left \{ \int_0^{k^+-k_1^+} dl^+
  \frac{l^+-k^++\frac{3}{2}k_1^+}{(k^+-k_1^+)^2} \frac{k_\perp ^2}{l_\perp^2}
-   \int_{k^+-k_1^+}^\infty \frac{dl^+ }{l^+}  \frac{k_\perp^2}{ l_\perp^2} \right \}
\end{eqnarray}
We rearrange the kinematic factor $k_\perp^2$ into the
soft part and combine it with the gluon TMD matrix element by partial integration,
\begin{eqnarray}
&& \!\!\!\!\!\!\!\!\!\!\!\!\!\!\!\!\!\!\!\! k_\perp^2 \
\langle  A^+_a(x^-\!,x_\perp)  g_0f^{bac}A^+_b(z^-\!,z_\perp)  A^+_c(y^-\!,y_\perp) \rangle
\nonumber \\
&& \Longrightarrow \langle  \partial_i A^+_a(x^-\!,x_\perp)  g_0f^{bac}A^+_b(z^-\!,z_\perp)
 \partial_i A^+_c(y^-\!,y_\perp) \rangle
\end{eqnarray}
Since the hard part is no longer dependent of $k_{1\perp}$, one can carry out the integration over $k_{1\perp}$.
This produces a delta function $\delta^2(z_\perp-y_\perp)$.
The integration for $z_\perp$ then can be trivially done. After performing the integral over $k_1^+$
by the residue theorem, one obtains,
\begin{eqnarray}
\frac{1}{2}\text{Fig.3(a)}+\text{Fig.3(c)} \approx
\frac{\alpha_s C_A}{4 \pi^2} \int \frac{d^2l_\perp}{l_\perp^2}
\left \{\int_{k^+}^\infty  \frac{dl^+}{l^+ }  +\frac{1}{2}
\right \}
xG_1(x,k_\perp)
\end{eqnarray}
which differs from the Fig.2(a) by a factor $1/2$.  We will show that another half contribution comes from
the combination $\frac{1}{2}\text{Fig.3(a)}+\text{Fig.3(b)}$.
To simplify the calculation of $\frac{1}{2}\text{Fig.3(a)}+\text{Fig.3(b)}$, we play the following
trick. One can treat the $k_1$ gluon as the collinear one(the error is power suppressed),
 and thus apply the Ward identity to the internal gluon line in Fig.3(b). The hard  part from Fig.3(b) can be
 subsequently separated into two parts,
\begin{eqnarray}
&& \!\!\!\!\!\!\!\!\!\!\!\!\!
\text{Fig.3(b)} \propto
{\cal H}_{b}(k,k_1)\int
\frac{dx^-d^2x_\perp}{(2\pi)^2} \frac{dy^-d^2 y_\perp}{(2\pi)^2}\frac{dz^-d^2 z_\perp}{(2\pi)^2}
\nonumber \\ && \!\!\!\!\!\!\!\!\! \times
e^{ik_\perp \! \cdot x_\perp}e^{-i(k_\perp\!-k_{1\perp}) \cdot y_\perp}e^{-ik_{1\perp } \! \cdot z_\perp}
e^{i(k^+\!-k_{1}^+)   y^-}\!e^{ik_{1}^+   z^-} \!
\langle  A^+_a(x^-\!,x_\perp)  g_0f^{bac}A^+_b(z^-\!,z_\perp)  A^+_c(y^-\!,y_\perp) \rangle_x
\end{eqnarray}
where ${\cal H}_{b}(k,k_1)$ reads,
\begin{eqnarray}
&&\!\!\!\!\!\! {\cal H}_{b}(k,k_1)\propto \frac{C_A}{2} \frac{\alpha_s}{4 \pi^3}
\int d^2k_{1\perp} d k_1^+
\int dl^+ d^2l_\perp dl^- \frac{-i}{k_1^+ + i \epsilon}
\nonumber \\
&& \!\!\!\! \times \!  \left \{ \!\!
\frac{[k_\perp \cdot(k_\perp-k_{1\perp}-l_\perp)][l_\perp \cdot (l_\perp-k_\perp+k_{1\perp})
] -l^-k^+ k_\perp \cdot l_\perp}
{\left[ 2(l^+ \!\!+\!k_1^+)l^-\!\! -\!(l_\perp\!\!+\!k_{1\perp})^2\!+\!i\epsilon \right ] \!
\left[ 2l^-(l^+\!\!-\!k^+\!+\!k^+_1)\!-\!(l_\perp\!\!-\!k_\perp\!+\!k_{1\perp})^2\! +\!i
\epsilon \right ]\!\left [l^-\!\!-\!i\epsilon \right ]\!\left [l^+\!+\!k_1^+\!+\!i\epsilon\right ]} \right .\
\nonumber \\&&  \left .\ -
\frac{[k_\perp \cdot(k_\perp\!-k_{1\perp}-l_\perp)][l_\perp \cdot (l_\perp\!-k_\perp+k_{1\perp})
] -l^-k^+ k_\perp \cdot l_\perp}{\left[ 2l^+l^-\! -l_\perp^2+i\epsilon \right ]\!
\left[ 2l^-(l^+\!-\!k^+\!+k^+_1)\!-\!(l_\perp\!-k_\perp\!+k_{1\perp})^2 \!+i
\epsilon \right ]\left [l^-\!-i\epsilon \right ]\left [l^+\!+k_1^+\!+\! i\epsilon\right ]} \right \}
\end{eqnarray}
Note that we relabeled the gluon momentum flow in the above equation. The internal gluon line
sandwiched by the two incoming gluon lines carries momentum $l$.
It is easy to check that the second term in the above equation is canceled out by the half of the contribution
from Fig.3(a). We are left with the first term once combing Fig.3(b) with the half of Fig.3(a),
\begin{eqnarray}
{\cal H}_{b+\frac{a}{2}}(k,k_1)& \propto&  \frac{C_A}{2} \frac{\alpha_s}{4 \pi^3}
\int d^2k_{1\perp} d k_1^+
\int dl^+ d^2l_\perp dl^- \frac{-i}{k_1^+ + i \epsilon}
 \frac{1}{ \left[ 2(l^+ +k_1^+)l^- -(l_\perp+k_{1\perp})^2+i\epsilon \right ]}
\nonumber \\&&\times
\frac{[k_\perp \cdot(k_\perp-k_{1\perp}-l_\perp)][l_\perp \cdot (l_\perp-k_\perp+k_{1\perp})
] -l^-k^+ k_\perp \cdot l_\perp}{\left[ 2l^-(l^+-k^++k^+_1)-(l_\perp-k_\perp+k_{1\perp})^2 +i
\epsilon \right ]\left [l^--i\epsilon \right ]\left [l^++k_1^++i\epsilon\right ]}
\end{eqnarray}
which can be further simplified by changing  integration variable $l \rightarrow l+k_1$ and neglecting
$k_{1\perp}$ in the numerator,
\begin{eqnarray}
{\cal H}_{b+\frac{a}{2}}(k,k_1)& \propto&  \frac{C_A}{2} \frac{\alpha_s}{4 \pi^3}
\int d^2k_{1\perp} d k_1^+
\int dl^+ d^2l_\perp dl^- \frac{-i}{k_1^+ + i \epsilon}
 \frac{1}{ \left[ 2l^+ l^- -l_\perp^2+i\epsilon \right ]}
\nonumber \\&\times&  \!
\frac{[k_\perp \cdot(k_\perp-l_\perp)][l_\perp \cdot (l_\perp-k_\perp)
] -l^-k^+ k_\perp \cdot l_\perp}{\left[ 2l^-(l^+-k^+)-(l_\perp-k_\perp)^2 +i
\epsilon \right ]\left [l^--i\epsilon \right ]\left [l^++i\epsilon\right ]}
\end{eqnarray}
This turns out to be the same as the hard part of Fig.2(a) except for the color factor.
Following the procedure outlined above, the UV pole term extracted from Fig.3(a), Fig.3(b) and Fig.3(c)
 is given by,
\begin{eqnarray}
\text{Fig.3(a)}\!+\!\text{Fig.3(b)}\!+\!\text{Fig.3(c)} \! &\approx& \!
\frac{\alpha_s C_A}{2\pi^2} \!\int \!\frac{d^2l_\perp}{l_\perp^2}
\left \{\int_{k^+}^\infty  \frac{dl^+}{l^+ } \! - \!\int_0^{k^+} \!\!dl^+
  \frac{l^+-k^+}{(k^+)^2}
\right \}
xG_1(x,k_\perp)
\nonumber \\&=&\frac{\alpha_s C_A}{2 \pi^2} \int \frac{d^2l_\perp}{l_\perp^2}
 \left \{ \frac{1}{2}+ \int_{k^+}^\infty \frac{dl^+}{l^+} \right \}xG_1(x,k_\perp)
\end{eqnarray}
The vertex correction now is correctly reproduced with one gluon re-scattering being taken into account.

Diagrams Fig.3(f) and Fig.3(g) also represent vertex correction, which however, do not contribute to the scale
evolution of gluon TMD. Instead, they are responsible for the running of the strong coupling constant in the
gauge link together with gluon self energy diagram Fig.3(i) and the Wilson line self energy diagram Fig.3(h).
The similar calculation for diagram Fig.3(f) leads to,
\begin{eqnarray}
{\cal H}_{f}(k_1,k_2) \!&\propto& \! \frac{C_A}{2} \frac{\alpha_s}{4 \pi^3} \int d^2k_{1\perp} d k_1^+
 \int \! dl^+ d^2l_\perp dl^- \frac{1}{k_1^++i \epsilon}
 \nonumber \\ && \times
\frac{ -i \left [l_\perp^2-k_{1\perp} \cdot l_\perp\right ]}{\left[ 2l^+l^- -l_\perp^2+i
\epsilon \right ]\left[ 2l^-(l^+-k^+_1)-(l_\perp-k_{1\perp})^2 +i
\epsilon \right ]\left [l^--i\epsilon \right ]\left [l^++i\epsilon
\right ]}
 \nonumber \\&=&\!
\frac{C_A}{2} \frac{\alpha_s}{2 \pi^2} \int d^2k_{1\perp}d^2l_\perp
 \int \!   d k_1^+\frac{-1}{k_1^++i \epsilon}
\nonumber \\&\times&\!\!
\left \{ \!\int_0^{k^+_1} \!\!\! dl^+   \frac{l_\perp^2-k_{1\perp} \cdot l_\perp}
{\left[ (l^+\!-k^+_1)l_\perp^2\!-(l_\perp\!-k_{1\perp})^2 l^+ \!+i\epsilon \right ]l_\perp^2}
-\!\int_{k^+_1}^\infty \!\! dl^+  \frac{l_\perp^2-k_{1\perp} \cdot l_\perp }
{(l_\perp\!-k_{1\perp})^2 l_\perp^2 l^+} \! \right \}
\end{eqnarray}
The external transverse momentum  $k_{1\perp}$ is set to be zero, one has,
\begin{eqnarray}
{\cal H}_{f}(k_1,k_2) \approx
\frac{C_A}{2} \frac{\alpha_s}{2 \pi^2} \int d^2k_{1\perp} \int \frac{d^2l_\perp }{l_\perp^2}
\int \frac{dk_1^+}{k_1^++ i\epsilon}
\left \{ \int_0^{k^+_1} \frac{dl^+}{k_1^+}
+ \int_{k^+_1}^\infty \frac{ dl^+ }{l^+}   \right \}
\end{eqnarray}
By carrying out the contour integration for  $k_1^+$,
the lower limit of the second integration is constrained to be zero. We arrive at,
\begin{eqnarray}
\text{Fig}.3(f) \approx
\frac{C_A}{2} \frac{\alpha_s}{2 \pi^2}  \int \frac{d^2l_\perp }{l_\perp^2}
  \left [1+\int_{0}^\infty \frac{ dl^+ }{l^+} \right ] xG_1(x,k_\perp)
\end{eqnarray}
The hard part of Fig.3(g) is written as,
\begin{eqnarray}
{\cal H}_g(k_1,k_2) &\propto & \frac{C_A}{2} \frac{\alpha_s}{4 \pi^3}
\int d^2k_{1\perp} d k_1^+  \int dl^+ d^2l_\perp dl^- \frac{1}{k_1^++i\epsilon}
\nonumber \\ &&\times
\frac{-i}{\left[ 2l^+l^- -l_\perp^2+i
\epsilon \right ]\left [l^--i\epsilon \right ]\left [k_1^++ l^++i\epsilon \right] }
 \nonumber \\&=&
  \frac{C_A}{2} \frac{\alpha_s}{2 \pi^2}\int d^2 k_{1\perp} \frac{d k_1^+}{k_1^+\!+i\epsilon}
   \int \frac{d^2l_\perp}{l_\perp^2} \int_{k_1^+}^{\infty} \frac{dl^+ }{l^++i\epsilon}
\end{eqnarray}
After integrating over $k_1^+$, one obtains,
\begin{eqnarray}
\text{Fig}.3(g) \approx
\frac{C_A}{2} \frac{\alpha_s}{2 \pi^2}  \int \frac{d^2l_\perp }{l_\perp^2}
  \int_{0}^\infty \frac{ dl^+ }{l^+}  xG_1(x,k_\perp)
\end{eqnarray}
The contributions from the Wilson line self energy are listed as the follows,
\begin{eqnarray}
 \text{Fig}.3(d)= \text{Fig}.3(h) =
 - \frac{\alpha_s C_A}{ 2\pi^2} \int \frac{d^2l_\perp}{l_\perp^2} \int_0^{\infty} \frac{dl^+}{l^+}
  xG_1(x,k_\perp)
\end{eqnarray}
and gluon self energy diagrams,
\begin{eqnarray}
 \text{Fig}.3(e) \approx \text{Fig}.3(i) \approx
 \frac{\alpha_s C_A}{ 2\pi^2} \int \frac{d^2l_\perp}{l_\perp^2} \frac{5}{12}
  xG_1(x,k_\perp)
\end{eqnarray}

We are now ready to assemble all pieces together.  First, one notices that
the light cone divergence is canceled out among Fig.3(f), Fig.3(g) and Fig.3(h). Including gluon
self energy diagram, we have,
\begin{eqnarray}
\text{Fig.3(f+g+h+i)}
 \approx
 \frac{\alpha_s C_A}{ 2\pi^2} \int \frac{d^2l_\perp}{l_\perp^2} \frac{11}{12}
  xG_1(x,k_\perp)
  \end{eqnarray}
 from which, one can reproduce the one loop beta function that describes the scale dependence of the strong
  coupling constant.
The summation of the rest diagrams in Fig.3 gives,
\begin{eqnarray}
{\text Fig}.3(a+b+c+d+e) &\approx& \frac{\alpha_s C_A}{2\pi^2}
\!\int \frac{d^{2} l_\perp}{l_\perp^2}\left \{\frac{11}{12}-\int_0^{k^+}
\frac{dl^+}{l^+} \right \}
\end{eqnarray}
Adding up hermitian  conjugate contributions and the soft factor,  the final result reads,
\begin{eqnarray}
 && \!\!\!\!\!\!\!\!\!\!\!\!\!\!\!\!\!\!\!\!\!\!\!\!\!\!\!
 \text{Fig}.3(a+b+c+d+e+f+g+h+i) \approx
 \nonumber \\&&\frac{\alpha_s C_A}{2\pi^2}
 (2\pi \mu)^{2\epsilon}
\!\int \!\frac{d^{2-2\epsilon} l_\perp}{l_\perp^2}\left \{\!\frac{11}{12}\!+\!\frac{11}{6}\!-
{\rm ln}\frac{\zeta_c^2}{l_\perp^2} \right \} xG_{ 1}(x,k_\perp)+ \!\text{UV c.t.}
\end{eqnarray}
The extra UV divergence can be removed by replacing the bare strong coupling constant $g_0$ with
 a renormalized one  $g$ in the gauge link,
\begin{eqnarray}
- \textbf{g}_0 \!\int_{y^-}^{-\infty}\!\!  dz^- f^{bac}A^+_b(z^-,y_\perp)
\Longrightarrow
- \textbf{g} \!\int_{y^-}^{-\infty}\!\!  dz^- f^{bac}A^+_b(z^-,y_\perp) \label{rep}
\end{eqnarray}
with
\begin{eqnarray}
g_0=g \mu^\epsilon \left [ 1-\frac{\alpha_s C_A  }{2\pi \epsilon}
S_\epsilon\frac{11}{12} \right ]
\end{eqnarray}
Here quark loop contribution is not included. The one loop virtual correction to the gluon TMD now takes form
\begin{eqnarray}
xG(x,k_\perp,\mu^2,\zeta_c^2) \! \approx \! \frac{\alpha_s C_A}{2\pi^2}
 (2\pi \mu)^{2\epsilon}
\!\int \!\frac{d^{2-2\epsilon} l_\perp}{l_\perp^2}\!\left \{\!\frac{11}{6}\!-\!
{\rm ln}\frac{\zeta_c^2}{l_\perp^2} \right \} \!\left \{ xG_0(x,k_\perp) \!+ \!xG_{\bar 1}(x,k_\perp)\right \}\!+ \!\text{UV c.t.}
\end{eqnarray}
where $xG_{\bar  1}$ denotes the gluon TMD with  the gauge link
 $- \textbf{g} \!\int_{y^-}^{-\infty}\!  dz^- f^{bac}A^+_b(z^-,y_\perp)$. It is easy to see that the
 anomalous dimension is not affected by gluon re-scattering effect.  This is more or less expected because
 the short distance physics(UV divergence) can  not be altered by physics happens in long distance(gluon re-scattering).

We now proceed to compute virtual correction with two gluon re-scattering. To generalize the
calculation to the two gluon re-scattering case, let us reexamine the evaluations of Fig.3(a),
Fig.3(b) and Fig.3(c) from a different aspect of view.  We start with investigating the $k_1^+$ pole structure
of these three diagrams. Fig.3(b) and Fig.3(c) generate the pole;
\begin{eqnarray}
\frac{1}{k_1^++l^++i\epsilon}
\end{eqnarray}
while a double pole emerges in Fig.3(a),
\begin{eqnarray}
\frac{1}{k_1^++l^++i\epsilon} \frac{1}{k_1^++i\epsilon}
\end{eqnarray}
If one picks up  $\frac{1}{k_1^++l^++i\epsilon}$ pole contributions and carries out the $l^-$ integration by
closing the contour circles around the pole $\frac{1}{l^2+i\epsilon} $, the Wilson line and the internal gluon
line are effectively put on shell. Moreover, the external transverse momenta can be neglected when computing
the UV counterterm. All exchanged gluons except for the left-most incoming gluon can be treated as the collinear ones.
The Ward identity argument then can be applied in a physical gauge calculation.
It is easy to see that the pole $\frac{1}{k_1^++l^++i\epsilon}$
contributions from three diagrams are canceled out due to the Ward identity.  We are left with the
 $\frac{1}{k_1^++i\epsilon}$ pole  contribution from Fig.3(a). At this step, one can directly isolate
the  $\frac{1}{k_1^++i\epsilon}$  pole contribution using the residue theorem. The rest calculation is
exactly same as that for the standard vertex correction represented by Fig.2(a).
\begin{figure}[t]
\begin{center}
\includegraphics[width=17 cm]{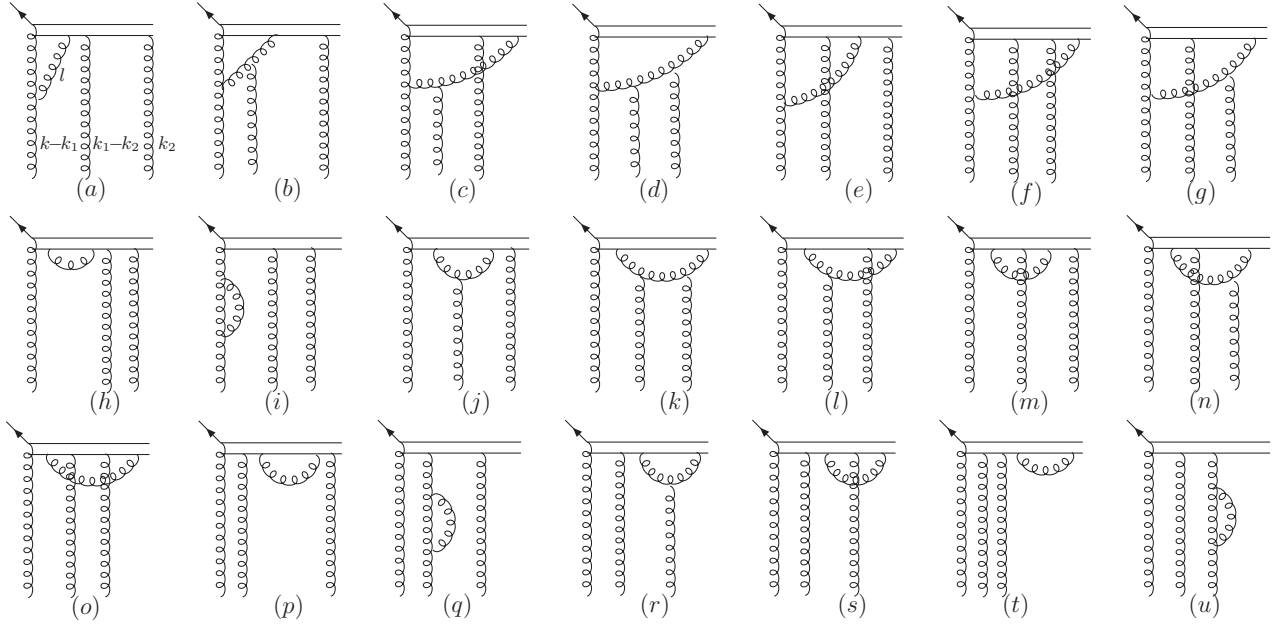}
\caption[] { One loop virtual correction with two gluon re-scattering.
Collinear reducible diagrams~\cite{collins} are not shown here. }
 \label{1}
\end{center}
\end{figure}

In an analogous way, the calculation of Fig.4 can be simplified by playing the same trick.
For instance, the $ \frac{1}{k_2^++l^++i\epsilon}$ pole contributions from Fig.4(j), Fig.4(k) and Fig.4(l)
are canceled out. Such cancelation also occurs among Fig.4(m), Fig.4(n) and Fig.4(o). We are left with
the $\frac{1}{k_2^++i\epsilon}$ pole contributions from Fig.4(j) and Fig.4(m). After carrying out integration
over $k_2^+$ using the residue theorem, it becomes obvious that the calculations of Fig.4(j) and Fig.4(m)
are the same as that for Fig.3(f) and Fig.3(g). We end up with,
\begin{eqnarray}
 \text{Fig}.4(j+k+l+m+n+o) \approx
\frac{\alpha_s C_A}{ 2\pi^2} \int \frac{d^2l_\perp}{l_\perp^2}
 \left [\frac{1}{2}+\int_0^\infty \frac{dl^+}{l^+} \right ]
  xG_2(x,k_\perp)
\end{eqnarray}
Other  vertex diagrams can be grouped and lead to,
\begin{eqnarray}
  \text{Fig}.4(r+s) \approx
 \frac{\alpha_s C_A}{ 2\pi^2} \int \frac{d^2l_\perp}{l_\perp^2}
 \left [\frac{1}{2}+\int_0^\infty \frac{dl^+}{l^+} \right ]
  xG_2(x,k_\perp)
\end{eqnarray}
Combining with the Wilson line self energy diagrams Fig.4(p) and Fig.4(t), and
gluon self energy diagrams Fig.4(q) and Fig.4(u), we obtain the UV pole contributions as,
\begin{eqnarray}
 \text{Fig}.4(i+j+k+l+m+n+o+p) \approx   \text{Fig}.4(r+s+t+u) \approx
\frac{\alpha_s C_A}{ 2\pi^2} \int \frac{d^2l_\perp}{l_\perp^2}
 \frac{11}{12}  xG_2(x,k_\perp)
 \nonumber \\
\end{eqnarray}
The UV divergence from these diagrams can be absorbed into the gauge link by making the following replacement,
\begin{eqnarray}
&& \textbf{g}_0^2 \!\int_{y^-}^{-\infty}\!\!\!\! dz^- f^{dab}A^+_d(z^-,y_\perp)
  \!\int_{y^-}^{z^-}\!\!\!\! dz_1^- f^{bdc}A^+_b(z_1^-,y_\perp)
  \nonumber \\
&  \Longrightarrow& \textbf{g}^2 \!\int_{y^-}^{-\infty}\!\!\!\! dz^- f^{dab}A^+_d(z^-,y_\perp)
  \!\int_{y^-}^{z^-}\!\!\!\! dz_1^- f^{bdc}A^+_b(z_1^-,y_\perp)
\end{eqnarray}

The rest diagrams contribute to the anomalous dimension of the gluon TMD.
Here we adopt the same strategy to simplify the calculation of the vertex correction
from Fig.4(a-g). One notices that Fig.4(b), Fig.4(c) and Fig.4(d) can be grouped together,
 while the cancelation of the $\frac{1}{k_2^++l^++i\epsilon }$ pole contributions occurs
 among Fig.4(e), Fig.4(f) and Fig.4(g).  After integrating out $k_2^+$, we
 are ready to take care of $k_1^+$ pole contributions in a similar manner. Eventually,
 the same vertex correction  is reproduced with graphs Fig.4(a-g).
 Combining with  the Wilson line self energy diagram Fig.4(h) and gluon self energy diagram Fig.4(i),
 one obtains,
\begin{eqnarray}
 \text{Fig}.4(a+b+c+d+e+f+g+h+i) \approx
\frac{\alpha_s C_A}{ 2\pi^2} \int \frac{d^2l_\perp}{l_\perp^2}
 \left [\frac{11}{12}-\int^{k^+}_0 \frac{dl^+}{l^+} \right ]
  xG_{ 2}(x,k_\perp)
\end{eqnarray}
 Now it is easy to see that the  identical UV pole structure of  the gluon TMD is
recovered  for the two gluon re-scattering case.

 The method introduced above can be recursively  applied to multiple gluon re-scattering case
 starting from the right-most gluon attachment. The evaluation of diagrams with $n$ gluon re-scattering
 always can be reduced to the calculation of diagrams with $n-1$ gluon attachment. The UV structure is not
 affected no matter how many soft gluons are exchanged. This is expected because as a quite general principle,
 long distance physics  does not cause any impact on short distance physics. We verified this statement
 by explicit calculations for this specific case.

Now let us summarize our calculation. We computed the gluon TMD defined in Eq.~\ref{TMDa} in term of
 the operator given in Eq.~\ref{TMDx}. The calculation is compatible with the conventional treatment of the small
 $x$ formalism.  To study the scale dependence of the gluon TMD, the UV pole terms
 from virtual corrections  are explicitly worked out. Apart from leading to the scale evolution of
  gluon TMD, virtual correction also results in other effects: 1) the running of strong coupling
constant, all the bare strong coupling constant in Eq.~\ref{TMDx}
should be replaced with the renormalized ones;
2) the bare gauge fields in Eq.~\ref{TMDx}  are replaced with the renormalzied fields.
The derived anomalous dimension of gluon TMD is
$\gamma_G \left(g(\mu),1\right )=\frac{\alpha_s C_A}{\pi} \frac{11}{6}$(it is straightforward to include
quark loop contribution), which is the same as the one calculated without taking into account multiple gluon re-scattering.
Therefore, we conclude that both the double leading and single leading logarithms can be resummed to all orders in saturation
regime by solving the CS evolution equation and the RG equation.

\section{summary}
This work is devoted to the study of the resummation of the single leading logarithms in saturation regime.
 In the Collins-2011 scheme, the double leading logarithm and single leading logarithm can be resummed into
 an exponentiation i.e. the Sudakov factor by solving the Collins-Soper equation and the RG equation.
 In a previous publication, we showed
 that small $x$ gluon TMDs do satisfy the Collins-Soper equation.  To derive the RG equation,
 we compute the one loop virtual corrections to the WW type  gluon TMD
 in the presence of multiple gluon re-scattering. As expected, the UV divergence structure of virtual
 corrections are not affected by multiple gluon re-scattering effect.  As a consequence, the anomalous
 dimension  of small $x$ gluon TMD determined through  UV pole terms is found to be
 the same as the one calculated  in the conventional way at one loop order.

Our analysis can be straightforwardly applied to other cases, for instance,
 the WW type gluon distribution with a future pointing gauge link and the dipole type gluon distribution.
 We reached the same conclusion that the perturbative part of
 the resulting Sudakov factor takes the same form in dense medium.
However, it is not yet clear if the non-perturbative part of the Sudakov factor is affected by saturation effect.
 We leave this for the future study.
 Nevertheless,  it is now clear that the full  $k_t$ resummation machinery can be employed to perform the
 relevant phenomenological studies of physical observables involve two well separated scales in saturation regime.

\

\

\noindent {\it \bf Acknowledgments:} I thank Feng Yuan and Bowen Xiao for helpful discussions.
This work has been supported by the National Science Foundation of China under Grant No. 11675093,
and by the Thousand Talents Plan for Young Professionals.

\end {document}